\documentclass[a4paper,12pt]{article}
\usepackage{amssymb,amsfonts,amsmath,mathtext,amsthm,cite,enumerate,float}

\usepackage{color}      % use if color is used in text

\newtheorem{proposition}{Proposition}
\newtheorem{theorem}{Theorem}
\newtheorem{corollary}{Corollary}
\newtheorem*{conjecture}{Conjecture}

\title{Estimating the output entropy of a tensor product of two quantum channels}

\author{Grigori G. Amosov\\
{Steklov Mathematical Institute}}

\begin{document}

\maketitle

\begin{abstract}
In this paper we find, for a class of bipartite quantum states, a nontrivial lower bound on the entropy gain 
resulting from the action of a tensor product of identity channel with
an arbitrary channel. By means of that we then estimate (from below) the output
entropy of the tensor product of \emph{dephasing} channel with an arbitrary channel.
Finally, we provide a characterization of all \emph{phase-damping} channels resulting 
as particular cases of \emph{dephasing} channels.
\end{abstract}

%\tableofcontents

\section {Introduction}

One of the most important tasks in quantum information theory is to calculate how informational capacities are changed 
under the external action on the quantum system.  Most of used capacities are determined on the base of the von Neumann entropy.
We consider the problem of finding a lower bound on the entropy gain for a state of bipartite system
in the case if the perturbation affects only its subsystem. The nontriviality  of this problem is
connected with a phenomenon of entanglement for bipartite quantum systems which doesn't exist for
classical systems.

Let us consider a state $\rho \in \mathfrak {S}(K)$, where 
 $\mathfrak {S}(K)$ denotes the set of all positive unit-trace operators  in
the Hilbert space $K$\footnote{Where not explicitly said, the Hilbert spaces are assumed to be infinite dimensional.}.
Then, given a quantum channel (a linear completely positive trace-preserving map) $\Omega $ on the set of
all bounded operators $B(K)$ on $K$, 
one can consider the entropy gain with respect to the action of a such channel
\begin{equation}\label{eg}
S(\Omega (\rho ))-S(\rho ).
\end{equation}
Here $S$ stands for the von Neumann entropy, $S(\rho )=-Tr(\rho \log \rho)$ with $0\le S(\rho)\le +\infty $.
The entropy gain is relevant since it measures the mixing property of the quantum channel \cite{Alicki}.

Let us consider the Kraus decomposition of $\Omega $
\begin{equation}\label{2}
\Omega (\rho )=\sum \limits _{j=1}^{+\infty}V_j\rho V_j^*,
\end{equation}
then if there exists
$$
s-\lim\limits _{N\to +\infty}\sum \limits _{j=1}^{N}V_jV_j^*=\Omega (I),
$$
and if $S(\rho)<+\infty$, 
the following lower bound for the entropy gain (\ref {eg}) has been found \cite {Hol}
\begin{equation}\label{Hol}
S(\Omega (\rho ))-S(\rho )\ge -Tr(\rho \log \Omega (I))\ge 0.
\end{equation}

%%%

Moving to the context of bipartite systems, 
it is of interest to evaluate the entropy gain for a state $\rho \in \mathfrak {S}(H\otimes K)$ under the action of the
tensor product of identity channel $Id :B(H)\to B(H)$ and the channel $\Omega$, i.e. 
\begin{equation}\label {holevo1}
S((Id \otimes \Omega )(\rho))-S(\rho ).
\end{equation}

Let us consider the convex closure $S_{\Omega }$ of the output
entropy for the state $\rho\in\mathfrak {S}(K)$ defined in \cite {HS} as 
\begin{equation}\label{holevo2}
S_{\Omega }(\rho)=\inf \limits _{\rho =\sum \pi _j\rho _j}\sum \limits _j\pi _jS(\Omega (\rho _j)).
\end{equation}
We conjecture the following relation between the two quantities (\ref {holevo1}) and (\ref {holevo2})
\begin{conjecture}
\begin{equation}\label{conj}
S((Id \otimes \Omega )(\rho))-S(\rho )\ge S_{\Omega }(Tr_H(\rho )).
\end{equation}
\end{conjecture}

In this paper we shall find a sufficiently broad class of states $\rho $ for which (\ref {conj}) holds true. 

\medskip

It is known that if the inequality (\ref {conj}) is valid for all states of 
the form $\rho =(\Phi \otimes Id)({\sigma} ),\ \sigma \in \mathfrak {S}(H\otimes K)$, 
then the minimal output entropy
would be additive with respect to tensor product of $\Phi $. 
Proving this property was one of the motivation to also address the Conjecture (see \cite {Rus}). It should be noticed that
states of the form $(\Phi \otimes Id)(|e><e|)$ are studied in \cite{DJKR} in the other context.

The quantum channel $\Phi$ is said to be {\it dephasing} \cite{SD, H} if there exists the orthonormal basis $(e_n)$ in $H$
such that
\begin{equation}\label{dp}
\Phi (|e_n><e_m|)=\lambda _{nm}|e_n><e_m|,
\end{equation}
where $(\lambda _{nm})$ is a positive definite matrix with $\lambda _{nn}=1$.
Quantum dephasing channels are know to be complementary to entanglement-breaking channels \cite{H}.

We shall show that (\ref {conj}) is related to the following inequality
\begin{equation}\label{conj2}
S((\Phi \otimes \Omega )(|e><e|))\ge S((\Phi \otimes Id)(|e><e|))+\sum \limits _{n}\pi _{n}S(\Omega (|h_n><h_n|)),
\end{equation}
where
$$
|e><e|=\sum \limits _{n,m}\lambda _n\overline \lambda _m|e_n><e_m|\otimes |h_n><h_m|,
$$
with orthonormal system $(e_n)$ in $H$, unit vectors $h_n\in K$,
$\Phi $ is  a dephasing channel while $\Omega $ is an arbitrary channel.
The inequality (\ref {conj2}) is closely related to the property of the strong superadditivity for the channel 
$\Phi $ introduced in \cite {HS} and widely discussed in \cite{AM}.

If $H=L^2({\mathbb R})$ the notion of a dephasing channel can be extended
to the channel $\Phi$ for which the output state $\Phi (|\psi ><\psi |)$ is the integral operator of the form \cite {H2}
\begin{equation}\label{g}
(\Phi(|\psi ><\psi | )\phi )(x)=\int \limits _{\mathbb R}\lambda (x,y)\psi (x)\overline \psi (y)\phi (y)dy,
\end{equation}
where $\phi \in L^2({\mathbb R})$ and $\lambda (x,y)$ is a positive definite kernel. We shall call (\ref {g}) {\it a generalised
dephasing} channel. 
Using the generalised eigenvectors of the position operator we can represent the action of (\ref {g})
in the form
\begin{equation}\label {g2}
\Phi (|x><y|)=\lambda (x,y)|x><y|.
\end{equation}
Equation (\ref {g2}) should be understood in the sense of (\ref {g}).

Let us consider a pure state
\begin{equation}\label{gr}
|e><e|=\int \limits _{\mathbb R^2}\lambda (x)\overline \lambda (y)|x><y|\otimes |h_x><h_y|dxdy,
\end{equation}
where $\int \limits _{\mathbb R}|\lambda (x)|^2dx=1$, $x\to h_x$ is a measurable function and $h_x$ are  unit vectors  in $K$.
More precisely (\ref {gr}) means that for a scalar product
of the bipartite system
$$
<\phi _1\otimes \psi _1|e><e|\phi _2\otimes \psi _2>=\int \limits _{{\mathbb R}^2}\lambda (x)\overline \lambda (y)\overline \phi _1(x)\phi _2(y)<\psi _1|h_x><h_y|\psi _2>dxdy,
$$
where $\phi _1,\phi _2\in H$ and $\psi _1,\psi _2\in K$.
Similarly to (\ref {conj2}) we shall show that
the following inequality holds true
\begin{equation}\label{conj3}
S((\Phi \otimes \Omega )(|e><e|))\ge S((\Phi \otimes Id)(|e><e|))+\int \limits _{\mathbb R}\pi (x)S(\Omega (|h_x><h_x|))dx
\end{equation}
with
$
\pi (x)=|\lambda (x)|^2,
$
where $\Phi $  a generalised dephasing channel and $\Omega $ an arbitrary channel.
 
The quantum dephasing channel (\ref {dp}) is said to be {\it a phase-damping} channel if
\begin{equation}\label {eq}
\lambda _{nm}=\lambda _{n-m},\ \overline {\lambda _n}=\lambda _{-n},
\end{equation}
for some complex numbers $(\lambda _n)_{n=0}^{N-1}$ which are the discrete Fourier transform of
a probability distribution $(\pi _n)_{n=0}^{N-1}$ \cite {Amo}.
In \cite {Amo} the inequality (\ref {conj2}) was obtained for 
a phase-damping channel $\Phi$ in the case $dimH<+\infty $.
We shall extend the definition of the phase-damping channel to the infinite-dimensional space.
Then, based upon Bochner's theorem \cite{Rudin} we shall give a complete classification of all phase-damping channels.

The paper is organized as follows. In Section 2 some results about the entropy gain will be proved. 
As an application the inequalities of the form (\ref {conj2}) and (\ref {conj3}) are 
obtained for all dephasing channels. Section 3 is devoted
to the description of quantum phase-damping channels of the form (\ref {eq}) as well as to their generalizations.
The last Section contains concluding remarks.

%%%%%%%%%%%%%%%%%%%%%%%%%%%%%%%%%%%%%%%

\section{The entropy gain}

We first derive a tighter bound on (\ref{Hol}) where the role of the identity operator $I$ is played by the orthogonal
projection $P$ with the property
$supp\rho \subset suppP$.

\begin{proposition}
Suppose that for $\rho \in \mathfrak {S}(H),\ S(\rho)<+\infty$, there is the orthogonal projection $P$
such that
$$
supp\rho \subset suppP,
$$
and the strong limit
$$
s-\lim \limits _{N\to \infty }\sum \limits _{j=1}^NV_jPV_j^*=\Omega (P),
$$
exists.
Then, it is
$$
S(\Omega (\rho ))-S(\rho )\ge -Tr(\Omega (\rho )\log \Omega (P)).
$$
\end{proposition}

%%%%%%%%%%%%%%%%%%%%%%%%%%%%%%

\proof{
We shall follow the techniques of \cite {Hol}.
Let us pick up an orthonormal basis $(e_k)$ spanning $suppP$ such that the state $\rho ,\ supp\rho \subset suppP,$
can be represented in the form
$$
\rho =\sum \limits _k\nu _k|e_k><e_k|.
$$
Under the condition of Proposition 1,
$$
S(\rho )=-Tr(\rho \log \rho)=\sum \limits _k\nu _k(-\log\nu_k)<+\infty .
$$
Then, there exists an operator $F$ satisfying
\begin{equation}\label{dop1}
Tr(\rho F)<+\infty ,\ Tr(\exp (-\beta F))<+\infty ,\ \beta > 0.
\end{equation}
Indeed, it suffices to set
$$
F=\sum \limits _{k}\mu _k(-\log \nu _k)|e_k><e_k|,
$$
where $(\mu _k)$ are taken in such a way that
$\mu _k\ \uparrow +\infty $ but $\sum \limits _k\mu _k\nu _k(-\log \nu_k)<+\infty $ still converges.
This allows us to define a state $\rho _{\beta }$ as follows
\begin{equation}\label{dop3}
\rho _{\beta }=\frac {P\exp (-\beta F)}{Tr(P\exp (-\beta F))}.
\end{equation}
The monotonicity of the quantum relative entropy $S(\rho\ ||\ \sigma )=Tr(\rho (\log\rho -\log\sigma ))$ gives
\begin{equation}\label{dop2}
S(\Omega (\rho )\ ||\ \Omega (\rho _{\beta }))\le S(\rho \ ||\ \rho _{\beta }).
\end{equation}
On the other hand,
$$
S(\rho \ ||\ \rho _{\beta })=-S(\rho)+\beta Tr(\rho F)+\log Tr(P\exp(-\beta F)),
$$
then (\ref {dop2}) implies that
\begin{equation}\label{dop4}
Tr(\Omega (\rho )(-\log \Omega (\rho _{\beta })))\le S(\Omega (\rho ))-S(\rho )+\beta Tr(\rho F)+\log Tr(P\exp(-\beta F)).
\end{equation}
Substituting (\ref {dop3}) in (\ref {dop4}) we get
$$
S(\Omega (\rho))-S(\rho )\ge Tr(\Omega (\rho )(-\log \Omega (P\exp (-\beta F))))-\beta Tr(\rho F).
$$
Since $P\exp (-\beta F)\le P$ it implies $\log \Phi (P\exp (-\beta F))\le \log \Phi (P)$ by the operator
monotonicity of the function $\log x$ on ${\mathbb R}_+$. Taking the limit $\beta \to 0$ we obtain the desired result. 
}

%%%%%%%%%%%%%%%%%%%%

We now move on by considering a bipartite system $H\otimes K$ where 
Proposition 1 allows us to prove the following theorem.

\begin{theorem} 
Suppose that $\rho \in \mathfrak {S}(H\otimes K)$ has the form
\begin{equation}\label{th2}
\rho =\sum \limits _{n,m}\lambda _{mn}|e_n><e_m|\otimes |h_n><h_m|,
\end{equation}
with orthonormal basis $(e_n)$ in $H$, unit vectors $h_n\in K$ and a positive definite matrix $(\lambda _{mn})$.
Then,
$$
S((Id\otimes \Omega )(\rho ))\ge S(\rho)+\sum \limits _n\pi _nS(\Omega (|h_n><h_n|)),
$$
where
$$
\pi _n=\lambda _{nn},
$$
and $\Omega $ is an arbitrary quantum channel.
\end{theorem}

\proof{
Suppose that a state $\rho \in \mathfrak {S}(H\otimes K)$ has the form (\ref {th2}). 
Following \cite {Amo} let us define
the orthogonal projection $P$ by the formula
\begin{equation}\label{proj}
P=\sum \limits _n|e_n><e_n|\otimes |h_n><h_n|.
\end{equation}
As consequence
$$
P\rho =\rho P=\rho,
$$
and hence
$$
supp\rho \subset suppP.
$$
Applying Proposition 1 to the state $\rho $ and the channel $\Phi =Id\otimes \Omega $ we get
$$
S((Id\otimes \Omega )(\rho))-S(\rho )\ge -Tr((Id\otimes \Omega )(\rho )\log (Id\otimes \Omega )(P)).
$$
Then using (\ref {proj}) the r.h.s. above results
$$
Tr((Id\otimes \Omega )(\rho )\log (Id\otimes \Omega )(P))=
$$
$$
Tr(\sum \limits _{n,m}\lambda _{nm}|e_n><e_m|\otimes \Omega (|h_n><h_m|)\sum \limits _k
|e_k><e_k|\otimes \log (\Omega (|h_k><h_k|)))=
$$
$$
Tr(\sum \limits _n\lambda _{nn}\Omega (|h_n><h_n|) \log (\Omega (|h_n><h_n|))).
$$
}

%%%%%%%%%%%%%%%%%%%%%%%

\begin{corollary}
Suppose that $\rho \in \mathfrak {S}(H\otimes K)$ has the form
\begin{equation}\label{th2}
\rho =\sum \limits _{n,m}\lambda _{mn}|e_n><e_m|\otimes |h_n><h_m|,
\end{equation}
with orthonormal basis $(e_n)$ in $H$, unit vectors $h_n\in K$ and a positive definite matrix $(\lambda _{mn})$.
Then,
$$
S((Id\otimes \Omega )(\rho ))-S(\rho)\ge S_{\Omega }(Tr_H(\rho )).
$$
\end{corollary}

\proof{
It immediately follows from the inequality
$$
\sum \limits _n\pi _nS(\Omega (|h_n><h_n|))\ge S_{\Omega }(\sigma),
$$
where $\sigma =\sum \limits _n\pi _n|h_n><h_n|$. 
}

%%%%%%%%%%%%%%%%%%%%%%%

\begin{corollary}
The relation (\ref {conj2}) holds for the dephasing channel $\Phi $ acting as
$$
\Phi (|e_n><e_m|)=\lambda _{nm}|e_n><e_m|.
$$
\end{corollary}

\proof{
Given a unit vector $e\in H\otimes K$ and an orthonormal basis $(e_n)$ in $H$ there exist the unit vectors $h_n\in K$
and the complex numbers $\nu _n,\ \sum \limits _n|\nu _n|^2=1,$ such that
$$
|e><e|=\sum \limits _{n,m}\nu _n\overline \nu _m|e_n><e_m|\otimes |h_n><h_m|.
$$
It results in the state
$$
\rho =(\Phi \otimes Id)(|e><e|)=\sum \limits _{n,m}\lambda _{nm}
\nu _n\overline \nu _m|e_n><e_m|\otimes |h_n><h_m|,
$$
satisfying the conditions of Theorem 1 if we replace $\lambda _{nm}$
by $\lambda _{nm}\nu _n\overline \nu _m$.
Then the result follows. 
}

%%%%%%%%%%%%%%%%%%%%%%%

Now, let us set $H=L^2(\mathbb {R})$ and consider the generalized eigenvectors $|x>$ of the position operator
$\hat x$ acting on $H$ as
$$
(\hat x f)(x)=xf(x),
$$
with $\cdot f(\cdot )\in H$, such that
\begin{equation}\label{x}
\hat x|x>=x|x>,
\end{equation}
$x\in \mathbb {R}$.
Applying Proposition 1 we shall also prove the following statement.

\begin{theorem}
Suppose that  $\rho \in \mathfrak {S}(H\otimes K)$ has the form
\begin{equation}\label{th3}
\rho =\int \limits _{{\mathbb R}^2}\lambda (x,y)|x><y|\otimes |h_x><h_y|dxdy,
\end{equation}
with unit vectors $h_x\in K$ and a positive definite function $\lambda (x,y)$.
Then,
$$
S((Id\otimes \Omega )(\rho ))\ge S(\rho)+\int \limits _{\mathbb R}\pi (x)S(\Omega (|h_x><h_x|))dx,
$$
where
$$
\pi (x)=\lambda (x,x),
$$
and $\Omega $ is an arbitrary quantum channel.
\end{theorem}

\proof{
Suppose that a state $\rho \in \mathfrak {S}(L^2({\mathbb R})\otimes K)$ has the form (\ref {th3}). Let us define
an orthogonal projection $P$ by the formula
\begin{equation}\label{proj2}
P=\int \limits _{\mathbb R}|x><x|\otimes |h_x><h_x|dx.
\end{equation}
It is straightforward to check that
$$
\rho P=P\rho =\rho .
$$
Therefore,
$$
supp\rho \subset suppP.
$$
Applying Proposition 1 to the state $\rho $ and the channel $\Phi =Id\otimes \Omega $ we obtain
$$
S((Id\otimes \Omega )(\rho))-S(\rho )\ge -Tr((Id\otimes \Omega )(\rho )\log (Id\otimes \Omega )(P)).
$$
Notice that using (\ref {th3}) and (\ref {proj2}) the r.h.s. above results in
$$
Tr((Id\otimes \Omega )(\rho )\log (Id\otimes \Omega )(P))=
$$
$$
Tr\left (\int \limits _{{\mathbb R}^2}\lambda (x,y)|x><y|\otimes \Omega (|h_x><h_y|)dxdy\int \limits _{\mathbb R}
|z><z|\otimes \log (\Omega (|h_z><h_z|))dz\right )=
$$
$$
Tr\left (\int \limits _{\mathbb R}\lambda (x,x)\Omega (|h_x><h_x|) \log (\Omega (|h_x><h_x|))\right ).
$$
}

%%%%%%%%%%%%%%%%%%%%%%%%%

\begin{corollary}
Suppose that  $\rho \in \mathfrak {S}(H\otimes K)$ has the form
$$
\rho =\int \limits _{{\mathbb R}^2}\lambda (x,y)|x><y|\otimes |h_x><h_y|dxdy,
$$
with unit vectors $h_x\in K$ and a positive definite function $\lambda (x,y)$.
Then,
$$
S((Id\otimes \Omega )(\rho ))-S(\rho)\ge S_{\Omega }(Tr_H(\rho )),
$$
where
$\Omega $ is an arbitrary quantum channel.
\end{corollary}

\proof{
It immediately follows from the inequality
$$
\int \limits _{\mathbb R}\pi (x)S(\Omega (|h_x><h_x|))dx\ge S_{\Omega}(\sigma ),
$$
where $\sigma =\int \limits _{\mathbb R}\pi (x)|h_x><h_x|dx$. 
}

%%%%%%%%%%%%%%%%%%%%%%%%
\begin{corollary}
The relation (\ref {conj3}) is satisfied for the generalised dephasing channel $\Phi$ acting as
$$
\Phi (|x><y|)=\lambda (x,y)|x><y|,
$$
and an arbitrary channel $\Omega $.
\end{corollary}

\proof{
Given a unit vector $e\in L^2({\mathbb R})\otimes K$ there exists a measurable function $x\to h_x$ acting from
the real line $\mathbb R$ to unit vectors $h_x\in K$ and the function $\nu (x),\ \int \limits _{\mathbb R}|\nu (x)|^2dx=1$
such that
\begin{equation}\label {z1}
|e><e|=\int \limits _{{\mathbb R}^2}\nu (x)\overline \nu (y)|x><y|\otimes |h_x><h_y|dxdy.
\end{equation}
Applying the generalised dephasing channel $\Phi $ to (\ref {z1}) we get the state
\begin{equation}\label{z2}
\rho =(\Phi \otimes Id)(|e><e|)=\int \limits _{{\mathbb R}^2}\lambda (x,y)\nu (x)\overline \nu (y)|x><y|\otimes |h_x><h_y|dxdy.
\end{equation}
It satisfies the condition of Theorem 2 if one replace $\lambda (x,y)$ by
$\lambda (x,y)\nu (x)\overline \nu (y)$.
}

%%%%%%%%%%%%%%%%%%%%%%%%%%%%%%%%%%%%%%%%%%%%%%%%%%%%%%%

\section{Quantum phase-damping channels}

Recall that a quantum dephasing channel $\Phi $, defined by the formula (\ref {dp}),
 is said to be a phase damping channel if the condition (\ref{eq}) is satisfied.
Analogously, starting from a generalised dephasing channel $\Phi$ defined by (\ref {g2}), 
we can introduce a generalised phase damping channel if the following condition is satisfied
\begin{equation}\label{eqq}
\lambda (x,y)=\lambda (x-y),\ \lambda (-x)=\overline \lambda (x).
\end{equation}

Since the kernels $\lambda _{nm}$ in (\ref {dp}) and $\lambda (x,y)$ in (\ref {g2}) are positive definite,
then so are the functions $\lambda _n$ and $\lambda (x)$, that is
$$
\sum \limits _{n,m}\lambda _{n-m}c_n\overline c_m\ge 0
$$
and
$$
\sum \limits _{n,m}\lambda (x_n-x_m)c_n\overline c_m\ge 0
$$
for any choice of $c_n\in {\mathbb C}$ and $x_n\in {\mathbb R},\ 1\le n\le N<+\infty $.
Thus, to classify all phase-damping channels we can apply the following theorem.

\begin{theorem}[Bochner's theorem \cite{Rudin}] 
Suppose that $f$ is a positive definite function on a locally compact Abelian group $G$ normalised by the
condition $f(e)=1$. Then, there exists
a unique probability measure $\mu $ on the dual group $\hat G$ such that
$$
f(g)=\int \limits _{\hat G}<\hat h,g>d\mu (\hat h).
$$
\end{theorem}

We shall consider three possible cases corresponding to $G={\mathbb Z}_N,\ {\mathbb Z}$ and $\mathbb R$.

\subsection{Finite dimension}

Suppose that the Hilbert space $H$ has a finite dimension, $dimH=N<+\infty $.
According to (\ref{dp}) $\Phi (|e_n><e_m|)=\lambda _{m-n}|e_n><e_m|$,
where $0\le n,m<N$. Then the following theorem holds true.

\begin{theorem}
The complex numbers
$(\lambda _n)$ are the discrete Fourier transform of a probability distribution $(\pi _n)_{n=0}^{N-1}$
determined by the formula
$$
\lambda _n=\sum \limits _{k=0}^{N-1}\exp \left (\frac {2\pi nki}{N}\right )\pi _k.
$$
Moreover, there exists a unitary operator $U:H\to H$ and an orthonormal basis $(f_n)$ in $H$ such that
$$
Uf_n=f_{n+1\ mod\ N},
$$
$0\le n\le N-1,$ and
$$
\Phi (\rho )=\sum \limits _{n=0}^{N-1}\pi _nU^n\rho U^{*n},
$$
$\rho \in \mathfrak {S}(H)$.
\end{theorem}

\proof{
Here it is more convenient to give a direct proof without using Bochner's theorem.
Let us consider the operator $T$ acting in ${\mathbb C}^N$ by the formula
$$
(T \nu)_m=\sum \limits _{n=0}^{N-1}\lambda _{m-n}\nu _n,
$$
where $\nu =(\nu _0,\dots ,\nu _{N-1})\in {\mathbb C}^N$.
Since $\lambda _n$ is a positive definite function we get
\begin{equation}\label{vspo2}
(\nu ,T \nu)\ge 0.
\end{equation}
Taking into account the Parseval equality for the discrete Fourier transform we obtain that
(\ref {vspo2}) results in
$$
\sum \limits _{n,m=0}^{N-1}\pi _n|\nu _n|^2\ge 0,
$$
for any choice of complex numbers $(\nu_n)$.
It implies that $\pi _n\ge 0$. On the other hand,
$$
\sum \limits _{n=0}^{N-1}\pi _n=\lambda _0=1.
$$
Then consider the unitary operator $U$ acting in $H$ as follows
\begin{equation}\label{u}
U|e_n>=\exp\left (\frac {2\pi n i}{N}\right )|e_n>,
\end{equation}
$0\le n<N$.
Let us define the orthonormal basis $(f_n)$ in $H$ by the formula
$$
|f_n>=\sum \limits _{m=0}^{N-1}\exp\left (\frac {2\pi nmi}{N}\right )|e_m>.
$$
Applying the operator (\ref {u}) to vectors $(f_n)$ we get
$$
U|f_n>=\sum \limits _{m=0}^{N-1}\exp \left (\frac {2\pi (n+1)m i}{N}\right )|e_m>=|f_n>.
$$
}

\subsection {Infinite dimension}

Now let $dimH=+\infty$. Fix the orthonormal basis $(e_n)$ and consider a quantum phase damping channel 
$\Phi$ defined by (\ref{2}).

\begin{theorem}
The complex numbers $(\lambda _n)$ are
the Fourier transform of a probability measure $\mu$ on the unit circle $\mathbb T$ such that
$$
\lambda _n=\int \limits _{\mathbb T}\exp\left (2\pi nti\right )d\mu (t).
$$
Moreover, there exists a unitary representation $t\to U_t$ of the multiplicative group $\mathbb T$ in $H$ such that
$$
\Phi (\rho )=\int \limits _{\mathbb T}U_t\rho U_t^*d\mu (t),
$$
$\rho \in \mathfrak {S}(H)$.
\end{theorem}

\proof{
Due to Bochner's theorem there exists a probability measure $\mu$ such that
$$
\lambda _n=\int \limits _{\mathbb T}\exp\left (2\pi nti\right )d\mu (t).
$$
Let us define a unitary representation ${\mathbb T}\ni t\to U_t$ by the formula
$$
U_t|e_n>=e^{2\pi nti}|e_n>,
$$
$n\in {\mathbb Z}$. Consider a quantum channel $\tilde \Phi$ of the following form
$$
\tilde \Phi (\rho )=\int \limits _{\mathbb T}U_t\rho U_t^*d\mu (t),
$$
$\rho \in \mathfrak {S}(H)$.
It follows that
$$
\tilde \Phi (|e_n><e_m|)=\int \limits _{\mathbb T}e^{2\pi (n-m)ti}d\mu (t)=\lambda _{n-m}|e_n><e_m|.
$$
Hence $\tilde \Phi =\Phi $. 
}

\subsection{The generalised phase-damping channel}

Let us consider here the case $H=L^2({\mathbb R})$. 
There the generalised quantum phase damping channel is defined by the formula
\begin{equation}\label{mapa2}
\Phi (|x><y|)=\lambda (x-y)|x><y|.
\end{equation}

\begin{theorem}
The function $\lambda (x)$ in (\ref {mapa2}) is the Fourier transform of a probability
measure $\mu$ on the line $\mathbb R$ defined by the formula
$$
\lambda (x)=\int \limits _{\mathbb R}\exp (ixy)d\mu (y).
$$
Moreover, there exists a strong continuous one-parameter group of unitaries $t\to U_t,\ U_0=I,$ such that
$$
\Phi (\rho )=\int \limits _{\mathbb R}U_t\rho U_t^*d\mu (t),
$$
$\rho \in \mathfrak {S}(H)$.
\end{theorem}

\proof{
Due to Bochner's theorem there exists a probability measure $\mu$ on the line $\mathbb R$ such that
$$
\lambda (x)=\int \limits _{\mathbb R}\exp (ixy)d\mu (y).
$$
Let us define a strong continuous group of unitaries $(U_t)$ by the formula
$$
(U_t\psi )(x)=e^{itx}\psi (x),\ \psi \in H.
$$
Consider a quantum channel $\tilde \Phi $ of the following form
$$
\tilde \Phi (\rho )=\int \limits _{\mathbb R}U_t\rho U_t^*d\mu (t),
$$
$\rho \in \mathfrak {S}(H)$.
It follows that for $\psi ,\phi ,\xi \in H$ we get
$$
(\tilde \Phi (|\psi ><\phi |)\xi )(x)=\int \limits _{{\mathbb R}^2}e^{it(x-y)}\psi (x)\overline \phi (y)\xi (y)dyd\mu (t)=
\int \limits _{\mathbb R}\lambda (x-y)\psi (x)\overline \phi (y)\xi (y)dy.
$$
It implies that $\tilde \Phi =\Phi $.
}

\section{Conclusion}

We have derived a nontrivial lower bound 
on the entropy gain with respect to the action of an arbitrary quantum channel affecting only one part of the system, 
for a class of bipartite quantum states  (Corollary 1 and 3). Based on this result we have estimated (from below)
the output entropy for tensor product of the dephasing quantum channel and an arbitrary channel (Corollary 2 and 4). Finally, we have introduced a classification
of quantum phase-damping channels resulting as special cases of the dephasing channels (Theorem 4, 5 and 6).

\section{Acknowledgments} The author is grateful to all participants of the seminar "Quantum Probability, Statistics, Information" in Steklov
Mathematical Institute for useful discussions. Especially the author likes to thank A.S. Holevo and S. Mancini whose careful reading and remarks
allowed to improve the exposition. The work is supported by the grant of Russian Scientific Foundation (Project N 14-21-00162).


\begin{thebibliography}{11}

\bibitem{Alicki} R. Alicki, "Isotropic quantum spin channels and additivity questions", arXiv:quant-ph/0402080.


\bibitem{Amo} G.G. Amosov, "On estimating the output entropy of a tensor product of the quantum phase-damping channel with an arbitrary channel", {\it Problems of Information Transmission} 49:3 (2013) 224–231; arXiv:1301.2886

\bibitem{AM}
G.G. Amosov, S. Mancini, "The decreasing property of relative entropy and the strong superadditivity 
of quantum channels", {\it Quantum Information and Computation} 9:7  (2009) 594-609


\bibitem{SD} I. Devetak, P. Shor, "The capacity of a quantum channel for simultaneous transmission of classical and quantum information", arXiv:quant-ph/0311131


\bibitem{Hol} A.S. Holevo, "The entropy gain of infinite-dimensional quantum channels", {\it Doklady Mathematics} 82:2 (2010) 730-731;
arXiv:1003.5765

\bibitem{H} A.S. Holevo, "On complementary channels and the additivity problem", {\it Theory Probab. Appl.}, 51:1 (2007), 92–100; arXiv:quant-ph/0509101

\bibitem{H2} A.S. Holevo, "Entanglement-breaking channels in infinite dimensions", {\it Problems of Information Transmission} 44:3 (2008) 3-18

\bibitem{HS} A.S. Holevo, M.E. Shirokov, "On Shor's channel extension and constrained channels", {\it Commun. Math. Phys.}
249 (2004) 417-430; arXiv:quant-ph/0306196

\bibitem{Rudin} W. Rudin "Fourier Analysis on Groups", {\it Wiley, New York}, 1990

\bibitem{Rus} M.B. Ruskai, "Some open problems in quantum information theory", arXiv:0708.1902.

\bibitem{DJKR}  I. Devetak, M. Junge, C. King, M.B. Ruskai, "Multiplicativity of completely bounded p-norms implies a new additivity result"\ , {\it Commun. Math. Phys.} 266  (2006) 37-63







\end{thebibliography}
\end{document}